\documentclass[12pt]{article}
\usepackage{graphicx}
\def \b{{\cal B}}
\def \bea{\begin{eqnarray}}
\def \beq{\begin{equation}}
\def \bra#1{\left| #1 \right \rangle}
\def \eea{\end{eqnarray}}
\def \eeq{\end{equation}}
\def \ite{{\it et al.}}

\def \lra{\leftrightarrow}
\def \ok{\overline{K}^0}
\def \s{\sqrt{2}}
\def \st{\sqrt{3}}

\begin{document}
\renewcommand{\thetable}{\Roman{table}}
\baselineskip 18pt
\begin{flushright}
TECHNION-PH-2005-12\\
EFI-05-14\\
hep-ph/0509155 \\
September 2005 \\
\end{flushright}

\centerline{\bf SYMMETRY RELATIONS IN CHARMLESS $B \to PPP$ DECAYS
\footnote{To be published in Physical Review D}}
\bigskip
\centerline{Michael Gronau$^2$ and Jonathan L. Rosner$^3$}
\medskip
\centerline{\it $^2$Department of Physics, Technion--Israel Institute of
Technology,}
\centerline{\it Technion City, 32000 Haifa, Israel}
\medskip
\centerline{\it $^3$Enrico Fermi Institute and Department of Physics,
University of Chicago}
\centerline{\it 5640 South Ellis Avenue, Chicago, IL 60637, USA}
\bigskip
\centerline{\large Abstract}
\bigskip
Strangeness-changing decays of $B$ mesons to three-body final states
of pions and kaons are studied, assuming that they are dominated by a 
$\Delta I=0$ penguin amplitude with flavor structure $\bar b \to \bar s$. 
Numerous isospin relations for $B\to K\pi\pi$ and for underlying quasi-two-body 
decays are compared successfully with experiment, in some cases resolving 
ambiguities in fitting resonance parameters. The only exception is a somewhat 
small branching ratio noted in $B^0\to K^{*0}\pi^0$, interpreted in terms of 
destructive interference between a penguin amplitude and an enhanced
electroweak penguin contribution.
Relations for $B$ decays into three kaons are derived in terms 
of final states involving $K_S$ or $K_L$,  assuming that $\phi K$-subtracted 
decay amplitudes are symmetric in $K$ and $\bar K$, as has been observed 
experimentally.  Rates due to nonresonant backgrounds are studied using a
simple model, which may reduce discrete ambiguities in Dalitz plot analyses.
\bigskip

\leftline{PACS numbers:  13.25.Hw, 11.30.Cp, 12.15.Ji, 14.40.Nd}
\bigskip

\centerline{\bf I.  INTRODUCTION}
\bigskip

The decays of $B$ mesons to charmless three-body final states provide valuable
information about the pattern of CP violation, as in the time-dependent
studies of CP asymmetries in decays to CP-eigenstates consisting of three
neutral pseudoscalars~\cite{Gershon:2004tk}.  Data for $B^0\to K^+K^-K_S$ of
comparable statistical weight have been presented by the BaBar
\cite{Aubert:2004ta,Aubert:2005ja,Aubert:2005kd} and Belle
\cite{Abe:2003yt,Garmash:2003er,Chen:2005dr} Collaborations.  In analyzing
these data it is of interest to know the CP eigenvalue of the three-body final
state which depends on the $K^+K^-$ angular momentum. 
In Ref.\ \cite{Garmash:2003er} isospin symmetry was
utilized to relate the decays $B^+ \to K^+ K^0 \bar K^0$ (measured via
$B^+ \to K^+ K_S K_S$) and $B^0 \to K^+ K^- K^0$ in order to conclude that
the $K^+ K^-$ final state was dominated by even angular momenta.

The question of genuine three-body decays of the $B$ meson (in contrast to
quasi-two-body decays which involve resonances between two of the three
bodies) arises in part because of the need to parametrize CP violation in $B
\to K \bar K K$ and to understand nonresonant contributions arising in Dalitz
plot studies.  These contributions can be quite large, as measured in Dalitz
plot analyses of $B^+\to K^+K^-K^+$~\cite{Garmash:2004wa} and $B^0 \to K^+ K^-
K_S$~\cite{Aubert:2005kd}.  They seem to be less significant in comparison with
quasi-two-body final states in certain $B \to K \pi \pi$
decays~\cite{Aubert:2004fn,Group(HFAG):2005rb}. 

In the present paper we discuss conclusions that can be drawn regarding the
structure of the three-body final states for $B \to PPP$ strangeness-changing
decays, where $P$ denotes a light pseudoscalar meson.  We begin by noting some
relations due to isospin in the limit in which decays are dominated by a QCD
penguin amplitude with isospin-preserving flavor structure.
Smaller $\Delta I = 1$ tree and electroweak penguin amplitudes will be
neglected.
We also analyze amplitudes
for a nonresonant background using isospin symmetry and flavor SU(3). The
description of $B$ decays to a pair of charmless mesons in terms of flavor
SU(3) amplitudes \cite{Gronau:1994rj,Gronau:1995hn} has been able to
correlate decay rates and CP-violating asymmetries for a wide variety
of processes involving two light pseudoscalars $P$ \cite{Chiang:2004nm}
or one pseudoscalar and one vector ($V$) meson \cite{Chiang:2003pm}.

We restrict our treatment for the moment in several respects.  (1) We
consider only $|\Delta S| = 1$ transitions and assume them to be dominated
by a penguin amplitude with flavor structure $\bar b \to \bar s$.  (2)
We consider only final states involving pions and kaons, in order not to
have to contend with octet-singlet mixing questions or posssible additional
flavor-singlet penguin amplitudes \cite{DGR}.  (3) We do not consider $B_s$
mesons, since information on them has lagged considerably behind that on
$B^+$ and $B^0$.

Earlier treatments of $B \to PPP$ decays, including model-dependent hadronic
calculations of decay rates and CP asymmetries, may be found in
Refs.~\cite{Eilam:1995nz,Bajc:1998bs,Cheng:2002qu,Minkowski:2004xf}.
Several analyses using flavor SU(3) have been performed in
Refs.~\cite{Gronau:2003ep,Grossman:2003qp},
in order to obtain model-independent bounds on deviations from the dominance
of a single weak phase in $B$ decays  to three kaons.

Section II reviews what is known about $B \to K \pi \pi$ and $B \to K \bar K K$
decay rates, pointing out certain features of resonant and nonresonant
contributions. Numerous isospin relations are proven for $B\to K\pi\pi$ and for
corresponding quasi two-body decays and are tested in Section III.  Similar
relations hold for $B\to K\bar K K$.  Assuming symmetry under the interchange
of $K$ and $\bar K$ momenta, as observed in the data, we prove decay rate
relations for processes involving $K_S$ and $K_L$.  Section IV compares
nonresonant background amplitudes in $B \to K \pi \pi$ and $B \to K \bar K K$
processes using isospin symmetry and flavor SU(3) in a simple universal model.
Implications for Dalitz plot analyses are noted in Section V, while Section VI
summarizes, concluding with a few remarks about isospin-violating corrections
and direct CP asymmetries.
\bigskip

\centerline{\bf II.  EXPERIMENTAL STATUS}
\bigskip

\begin{table}
\caption{Summary of CP-averaged branching ratios, in units of $10^{-6}$, for
$B \to K \pi \pi$ and $B \to K \bar K K$ including quasi two-body 
decays~\cite{Group(HFAG):2005rb,Abe:2005kr}.
Pairs of processes in the same row are related by isospin reflection, 
except for $B^+\to \phi K^+$ and $B^0\to \phi K^0$ which are in different rows.
\label{tab:brs}}
\begin{center} 
\begin{tabular}{l c c c} \hline \hline
Final state & Branching ratio & Final state & Branching ratio \\
in $B^+$ decay & $\times 10^{-6}$ & in $B^0$ decay & $\times 10^{-6}$ \\ 
\hline \hline
$K^+ \pi^+ \pi^-$ & $54.1\pm3.1$ &  $K^0 \pi^+ \pi^-$ & $44.9\pm2.6$ \\
$K^{*0}\pi^+$ & $10.8 \pm 0.8$ & $K^{*+}\pi^-$ & $9.8 \pm 1.1$ \\
$K^+\rho^0$ & $4.23^{+0.56}_{-0.57}$ & $K^0\rho^0$ & $ 5.6 \pm 1.1$ \\
$K^+f_0(980)$ & $9.1^{+0.8}_{-1.1}~^a$ & $K^0f_0(980)$ & $6.0 \pm 0.9~^a$ \\
$K^*_0(1430)^0\pi^+$ & $38.2^{+4.6}_{-4.5}$ & $K^*_0(1430)^+\pi^-$ &
$45.1 \pm 6.1$ \\ \hline
$K^0 \pi^+ \pi^0$ &    $< 66$  &  $K^+ \pi^- \pi^0$ & $35.6^{+3.4}_{-3.3}$ \\
$K^{*+}\pi^0$ & $6.9 \pm 2.3$ & $K^{*0}\pi^0$ & $1.7 \pm 0.8$ \\
$K^0\rho^+$ & $ < 48$ &  $K^+\rho^-$ & $9.9^{+1.6}_{-1.5}$ \\
$K^*_0(1430)^+\pi^0$ & -- & $K^*_0(1430)^0\pi^0$ & $7.9 \pm 3.1~^b$ \\ \hline
$K^+ \pi^0 \pi^0$ & -- &  $K^0 \pi^0 \pi^0$ &  --  \\ \hline
$K_S K_S K^+$  & $11.5\pm1.3$ & $K^+ K^- K^0$   & $24.7\pm2.3$ \\
~& ~& $\phi K^0$ & $8.3^{+1.2}_{-1.0}$ \\ \hline
$K^+ K^- K^+$   & $30.1\pm1.9$  & $K_S K_S K_S$  & $6.2\pm0.9$ \\
$\phi K^+$ & $9.03^{+0.65}_{-0.63}$ & ~ &
\\ \hline \hline
\end{tabular}
\end{center}
\leftline{$^a$ Includes $\b(f_0(980) \to \pi^+\pi^-)$.}
\leftline{$^b$ Includes $\b(K^*_0(1430)^0 \to K^+\pi^-)$.} 

\end{table}

The current world averages of CP-averaged branching ratios from BaBar, Belle,
and CLEO for the decays $B \to K \pi \pi$ and $B \to K \bar K K$ are summarized
in Table \ref{tab:brs} \cite{Group(HFAG):2005rb}.  Averages involving $B^0 \to
K_S \pi^+ \pi^-$ and its sub-modes
include recent Belle results \cite{Abe:2005kr}.  Also listed
are branching ratios for quasi-two-body decays for several resonances
contributing to these decays. 
While no measurement exists so far for the branching ratio of  
$B^0\to K^0\pi^0\pi^0$,  a time-dependent CP asymmetry has been 
recently reported in this process~\cite{Aubert:2005id}.

The branching ratios quoted in Table \ref{tab:brs} for $B^{+,0}\to K^{+,0}
f_0(980)$
and $B^0\to K^*_0(1430)^0\pi^0$ include
decay branching ratios of the daughter scalar mesons into observed modes.
Using
$\b(K^*_0(1430)^0 \to K^+\pi^-) = 2/3$ we obtain
\beq
 \b(B^0\to K^*_0(1430)^0\pi^0) = 11.9 \pm 4.7~~,
 \eeq
where branching ratios here and subsequently are quoted in units of $10^{-6}$.

Dalitz plot analyses of $B^0 \to K^+ K^- K^0$~\cite{Aubert:2005kd} and 
$B^+ \to K^+ K^- K^+$~\cite{Garmash:2004wa} find large nonresonant 
contributions in these decays. In addition to the $\phi K$ 
mode, where the $K^+$ and $K^-$ are in a P-wave, two sizable and comparable 
contributions have been measured: A term peaking around $1500~{\rm MeV}/c^2$, 
for which one finds in addition to a large solution also a small 
solution~\cite{Aubert:2005kd,Garmash:2004wa} 
(see Table I in Ref.~\cite{Aubert:2005kd} and Table V in Ref.~\cite{Garmash:2004wa}), 
and a term spreading across phase space. Both terms have an S-wave behavior in 
the $K^+$ and $K^-$ momenta~\cite{Aubert:2005kd,Garmash:2004wa}. 
Contributions from higher waves were found consistent with zero.  The decays 
$B^{+,0}\to \chi_{c0}K^{+,0}$, also having an S-wave behavior, contribute 
about three percent of the total branching ratios of $B^{+,0}\to
K^+K^-K^{+,0}$~\cite{Aubert:2005kd,Garmash:2004wa}.  Ref.~\cite{Garmash:2004wa}
finds a second solution of about eight percent for the fraction corresponding
to $B^+\to \chi_{c0}K^+$
(see Table V in~\cite{Garmash:2004wa}).
All the above three S-wave contributions are symmetric under interchanging 
$K^+$ and $K^-$.

It is useful to subtract contributions for $B \to \phi K$ from the branching
ratios of $B^+ \to K^+ K^- K^+$ and $B^0 \to K^+K^- K^0$. Using values in Table
I and~\cite{Eidelman:2004wy} $\b(\phi\to K^+ K^-) = (49.1\pm0.6)\%$, 
one finds
\bea\label{B+}
\b(B^+\to K^+K^-K^+)_{\phi K-{\rm subtracted}} & = & 25.7 \pm 1.9~~,\\
\label{B0}
\b(B^0\to K^+K^-K^0)_{\phi K-{\rm subtracted}} & = & 20.6 \pm 2.4~~.
\eea  
An important feature of the amplitudes corresponding to these branching ratios 
is their symmetry with respect to interchanging $K^+$ and $K^-$ momenta, 
as they are superpositions of three S-wave 
contributions~\cite{Aubert:2005kd,Garmash:2004wa}.

\newpage

\centerline{\bf III.  ISOSPIN RELATIONS}
\bigskip

We assume that the dominant transition for $|\Delta S| = 1$ $B \to PPP$ decays
has a flavor structure $\bar b \to \bar s$, which is isospin-invariant
($\Delta I = 0$).  Using isospin reflection symmetry under $u \lra d$,
we then find that each $B^+$ decay amplitude listed in Table \ref{tab:brs}
is equal (up to a possible sign) to a corresponding $B^0$ decay amplitude
listed in the same line.  Other amplitude relations follow from our assumption
that the final state is dominantly $I=1/2$.
In order to relate predictions of equal $B^+$ and $B^0$ partial widths to
observed branching ratios, we use the measured ratio of $B^+$ and $B^0$
lifetimes, $\tau_+/\tau_0 = 1.076\pm 0.008$ \cite{Group(HFAG):2005rb}.

Relations between observed branching ratios for $B\to K\pi\pi$ and
corresponding quasi two-body decays follow directly from the above assumption.
Similar amplitude relations hold for $B\to K\bar KK$.  However, in order to
rewrite these relations for decay rates involving $K_S$ and $K_L$ in the final
state one must assume a given symmetry under interchanging
$K$ and $\bar K$ momenta. We will use the symmetry under $K^+ \leftrightarrow
K^-$ of the amplitudes describing the $\phi K$-subtracted branching ratios
(\ref{B+}) and (\ref{B0}).

\bigskip

\leftline{\bf A.  $B \to K \pi \pi$}
\bigskip

A relation which is well-satisfied is
\bea
\b(B^+ \to K^+ \pi^+ \pi^-)&=&(\tau_+/\tau_0)\b(B^0 \to K^0 \pi^+ \pi^-)~~; \\
54.1 \pm 3.1 &=& 48.3 \pm 2.8 ~~.\nonumber
\eea
The discrepancy is only $1.4 \sigma$.

The above isospin  relation should apply to corresponding quasi-two-body 
modes contributing to these decays. Thus, the following four relations
hold reasonably well:
\bea 
\b(B^+\to K^{*0}\pi^+) & = & (\tau_+/\tau_0)\b(B^0\to K^{*+}\pi^-)~~;\\
10.8 \pm 0.8 & = & 10.6 \pm 1.2~~,\nonumber
\eea
\bea
\b(B^+\to K^+\rho^0) & = & (\tau_+/\tau_0)\b(B^0\to K^0\rho^0)~~;\\
4.23^{+0.56}_{-0.57} & = & 6.1 \pm 1.2~~,\nonumber
\eea
\bea
\b(B^+\to K^+f_0(980))\b(f_0\to\pi^+\pi^-) & = & 
(\tau_+/\tau_0)\b(B^0\to K^0f_0(980))\b(f_0\to\pi^+\pi^-);\\
9.07^{+0.81}_{-1.06} & = & 6.4 \pm 0.9~~,\nonumber
\eea
\bea\label{1430-1}
\b(B^+\to K^*_0(1430)^0\pi^+)
 & = & (\tau_+/\tau_0)\b(B^0\to K^*_0(1430)^+\pi^-)~~;
\\
38.2^{+4.6}_{-4.5}  & = & 48.6 \pm 6.6~~.\nonumber
 \eea 
The last relation disfavors a second solution, $\b(B^+\to K^*_0(1430)^0\pi^+)=
8.7 \pm 2.3$ measured in~\cite{Garmash:2004wa} 
(see Table IV there), and is in agreement with a more recent measurement
(see Table V in~\cite{Abe:2005ig}), $\b(B^+\to K^*_0(1430)^0\pi^+) = 
51.6 \pm 1.7 \pm 6.8^{+1.8}_{-3.1}$.

A prediction satisfied only by an upper bound is
\bea
\b(B^+ \to K^0 \pi^+ \pi^0)
 &=& (\tau_+/\tau_0)\b(B^0 \to K^+ \pi^- \pi^0)~~; \\
< 66 & = & 38.3^{+3.7}_{-3.6}~~.\nonumber
\eea
It should not be too difficult to obtain a value for the left-hand side; a
$\pi^0$ must be added to the observed final state $B^+ \to K^0 \pi^+$.
Corresponding predictions apply to quasi two-body decays. The 
prediction
\bea\label{K*pi1}
\b(B^0 \to K^{*0}\pi^0) & = & (\tau_0/\tau_+)\b(B^+\to K^{*+}\pi^0)~~; \\ 
1.7\pm 0.8 & = & 6.4 \pm 2.1~~,\nonumber
\eea
requires more data for a statistically significant test. 

Dominance of $I=1/2$ in $K^*\pi$ final states implies
\bea\label{K*pi2}
2\b(B^0\to K^{*0}\pi^0) & = & \b(B^0\to K^{*+}\pi^-)~~;\\
3.4 \pm 1.6 & = & 9.8 \pm 1.1~~,\nonumber
\eea
which is violated by $3.3\sigma$. 
The smallness of the left-hand side may be due to its sensitivity to small
$\Delta I = 1$ contributions, which are present in the treatment of Ref.\
\cite{Chiang:2003pm}.
The small branching ratio measured for $B^0\to K^{*0}\pi^0$ is evidently the
origin of the apparent discrepancies in Eqs.~(\ref{K*pi1}) and (\ref{K*pi2}). 

The prediction 
\bea
\b(B^+\to K^0\rho^+) & = & (\tau_+/\tau_0)\b(B^0\to K^+\rho^-)~~;\\
< 48 & =  & 10.7^{+1.7}_{-1.6}~~,\nonumber
\eea
is satisfied by the upper bound,
while $I(K\rho)=1/2$ implies
\bea
2\b(B^0\to K^0\rho^0) & = & \b(B^0\to K^+\rho^-)~~; \\
    11.3 \pm 2.2 & = & 9.9^{+1.6}_{-1.5}~~.\nonumber
\eea

Similarly,
\beq
\b(B^+\to K^*_0(1430)^+\pi^0) = (\tau_+/\tau_0)\b(B^0\to K^*_0(1430)^0\pi^0) =
12.8 \pm 5.0
\eeq
awaits a measurement of the left-hand-side, while $I(K^*_0(1430)\pi) = 1/2$
implies
\bea\label{1430-2}
2\b(B^0\to K^*_0(1430)^0\pi^0) & = & \b(B^0\to K^*_0(1430)^+\pi^-)~~;\\
23.7 \pm 9.3 & = & 45.1 \pm 6.1~~.\nonumber
\eea

Finally, neither the left-hand nor right-hand side of the following prediction
corresponds to a current observation:
\beq
\b(B^+ \to K^+ \pi^0 \pi^0) = (\tau_+/\tau_0)\b(B^0 \to K^0 \pi^0 \pi^0)~~.
\eeq

\bigskip\bigskip

\leftline{\bf B.  $B \to K \bar K K$}
\bigskip

Isospin reflection symmetry implies
\bea\label{00+}
A(B^+\to K^0 \ok K^+) & = & -A(B^0\to K^+K^-K^0)~~,\\
\label{+-+}
A(B^+\to K^+K^-K^+) & = & -A(B^0\to K^0\ok K^0)~~.
\eea
In order to study $\phi K$-subtracted amplitudes, we will use their observed
symmetry under interchanging the $K$ and $\bar K$ 
momenta mentioned at the end of Section II~\cite{Aubert:2005kd,Garmash:2004wa}.  
This permits writing relations for rates involving $K_S$ and $K_L$ in the final state.  
Note that because of Bose symmetry the amplitudes in (\ref{+-+}), which involve two
identical $K$ mesons, are also symmetric in the two $K$ momenta.

Using the phase convention (we neglect a tiny CP violation in $K^0$--$\ok$ mixing),
\beq
K_S \equiv (K^0 + \ok)/\s~~,~~~~~~~K_L \equiv (K^0 - \ok)/\s~~,
\eeq
a symmetric state
\beq
\bra{K^0 \ok}_{\rm sym} \equiv \left[ \bra{K^0(p_1) \ok(p_2)}
                              + \bra{\ok(p_1) K^0(p_2)} \right]/ \s~~,
                              \eeq
can be expressed as
\beq\label{KKbarSym}
\bra{K^0 \ok}_{\rm sym} = (\bra{K_S(p_1) K_S(p_2)}
                         - \bra{K_L(p_1) K_L(p_2)})/\s~~,
\eeq
while an antisymmetric state is given by 
\bea
\bra{K^0 \ok}_{\rm anti} & \equiv & \left[ \bra{K^0(p_1) \ok(p_2)}
- \bra{\ok(p_1) K^0(p_2)} \right] \s\nonumber\\
& = &\left[ \bra{K_L(p_1) K_S(p_2)} - \bra{K_S(p_1) K_L(p_2)} \right] /\s~~.
\eea
Similarly, a state symmetric in the three momenta, $p_1, p_2, p_3$, is given by
\bea\label{3K}
& & \bra{K^0 \ok K^0}_{\rm sym} \equiv \nonumber \\
& &  \left[ \bra{K^0 K^0 \ok}
+ \bra{K^0 \ok K^0}  + \bra{\ok K^0 K^0}\right]/ \st \\
& & = \frac{1}{2\s}\left [ \st \bra{K_S K_S K_S} + 
\bra{K_S K_S K_L}_{\rm sym}
-  \bra{K_L K_L K_S}_{\rm sym}
- \st  \bra{K_L K_L K_L}\right ] \nonumber~~,
\eea
where dependence on the three momenta has been suppressed.

Using Eq.~(\ref{KKbarSym}), we find 
\bea
\b(B^+ \to (K^0 \ok)_{\rm sym}K^+) & = & \b(B^+ \to K_S K_S K^+) + 
\b(B^+ \to K_L K_L  K^+)\nonumber \\
& = &  2\b(B^+ \to K_S K_S K^+) = 23.0 \pm 2.6~~.
\eea 
Eq.~(\ref{00+}) is well-satisfied for the $\phi K$-subtracted branching  ratio
\bea\label{1KKK}
\b(B^+ \to (K^0 \ok)_{\rm sym}K^+) & = &  (\tau_+/\tau_0) \b(B^0 \to
K^+K^-K^0)_{\phi K-{\rm subtracted}}~~;\\
23.0 \pm 2.6 & = & 22.2 \pm 2.6~~.\nonumber
\eea

Finally, in order to test the prediction (\ref{+-+}) we apply (\ref{3K}) 
\beq
\b(B^0 \to (K^0\ok K^0)_{\rm sym}) = \frac{8}{3}\b(B^0 \to K_SK_SK_S)
 = 16.5 \pm 2.4~~.
\eeq
Eq.~(\ref{+-+}) then reads
\bea\label{2KKK}
\b(B^+\to K^+K^-K^+)_{\phi K-{\rm subtracted}} & = & (\tau_+/\tau_0) \b(B^0\to 
(K^0 \ok K^0)_{\rm sym})~~;
\\
25.7 \pm 1.9 & = & 17.8 \pm 2.6~~,\nonumber
\eea
which holds within 2.4$\sigma$.
A potential discrepancy may be accounted for by small $\Delta I=1$ amplitudes.

Eq.~(\ref{3K}) also implies predictions for branching ratios involving $K_L$ in
the final state:
\bea
\b(B^0\to K_L K_L K_L) & = & \b(B^0 \to K_S K_S K_S) = \nonumber\\
3\b(B^0\to K_L K_L K_S) & = & 3\b(B^0\to K_S K_S K_L)~~,
\eea
where subtraction of $\phi K$ contributions in the last two processes is
implied.  We expect $\b(B^0\to K_S K_S K_L)$ to be easier to measure in
comparison with $\b(B^0\to K_L K_L K_S)$ and $\b(B^0\to K_L K_L K_L)$. 

The agreement in (\ref{1KKK}) and (\ref{2KKK}) supports the initial 
suggestion~\cite{Garmash:2003er} that the $K^+ $ and $K^-$ in the respective 
processes are in dominantly symmetric even angular momentum (S-wave) states, 
as confirmed directly by measuring angular dependence in later experiments
performing full Dalitz plot analyses~\cite{Aubert:2005kd,Garmash:2004wa}.
A statistically significant discrepancy in Eq.~(\ref{2KKK}), implied by reduced
experimental errors, would provide evidence for nonzero contributions either 
from odd angular momentum states or from a $\Delta I=1$ amplitude.
\bigskip

\centerline{\bf IV.  MODEL FOR A NONRESONANT BACKGROUND}
\bigskip
The measured $\phi K$-subtracted rates for $B^+\to K^+K^-K^+$
\cite{Garmash:2004wa} and $B^0 \to K^+K^-K^0$~\cite{Aubert:2005kd} consist each
of a sum of three contributions, all symmetric in the $K^+$ and $K^-$ momenta:
A small $\chi_{c0} K$ term, an S-wave contribution peaking around 1500
MeV/$c^2$, and a nonresonant background amplitude also representing an S-wave
in $K^+K^-$. The latter amplitude shows no significant dependence on the
$K^+K^+$ and $K^+K^0$ invariant masses in the two processes.  Some dependence
on the $K^+K^-$ invariant mass  is observed in $B^0\to K^+K^-K^0$ but not in 
$B^+\to K^+K^-K^+$. In a similar analysis of the nonresonant background in 
$B^+\to K^+\pi^+\pi^-$ some dependence  was measured on the invariant masses of 
$K^+\pi^-$ and of $\pi^+\pi^-$.

\begin{table}
\caption{Branching ratios of nonresonant background (NRB) contributions
for $B\to K\pi\pi$ and $B\to K\bar K K$, given as fractions 
of total branching ratios and in units of $10^{-6}$.  
\label{tab:NRB}}
\begin{center} 
\begin{tabular}{l ccc} \hline \hline
Decay mode & NRB fraction ($\%$) & NRB branching ratio
 ($\times 10^{-6}$) \\ \hline \hline
$B^+\to K^+\pi^-\pi^+$ &  $34.0 \pm 2.2^{+2.1}_{-1.8}$~\cite{Abe:2005ig}  & 
$18.4 \pm 1.9$~$^a$ \\
$B^0 \to K^+\pi^-\pi^0$ & ~ & $<4.6$~\cite{Group(HFAG):2005rb} \\
$B^+\to K^+K^-K^+$ & $74.8 \pm 3.6$~$^b$~\cite{Garmash:2004wa}
 & $22.5 \pm 1.8$ \\
$B^0 \to K^+K^-K^0$ & $70.7 \pm 3.8 \pm 1.7$~\cite{Aubert:2005kd}
 & $17.5 \pm 1.9$ \\
\hline \hline
\end{tabular}
\end{center}
\leftline{$^a$ A much smaller nonresonant branching ratio,
$(2.9^{+1.1}_{-0.9})\times 10^{-6}$, 
is quoted in~\cite{Group(HFAG):2005rb} .}
\leftline{$^b$ A second solution, $(65.1 \pm 5.1)\%$, is obtained
in~\cite{Garmash:2004wa}.}
\end{table}

In the present section we will study nonresonant background amplitudes in 
$B\to K\pi\pi$ and $B\to K\bar K K$ decays, adopting a simplified assumption 
that these amplitudes are symmetric under interchanging the three final meson
momenta.  This would be the case, for instance if the nonresonant amplitudes
were constant over the Dalitz plane; however, these amplitudes do not have to
be constant.  We start by first presenting the data and
then discussing symmetry relations governing nonresonant contributions.

Table \ref{tab:NRB} quotes measured fractions of nonresonant background (NRB) 
contributions in $B\to K\pi\pi$ and $B\to K\bar KK$ processses.  While a small
NRB contribution, $\b({\rm NRB})=(2.9^{+1.1}_{-0.9})\times 10^{-6}$, has been
measured in~\cite{Aubert:2004fn} and is quoted in~\cite{Group(HFAG):2005rb},
we quote in the Table a larger nonresonant fraction ($\sim 1/3$) which has been
measured recently by Belle~\cite{Abe:2005ig}
(see also Ref.~\cite{Garmash:2004wa}). As we will see, a large
nonresonant background in $B^+\to K^+\pi^+\pi^-$ appears to be more consistent
in our scheme with comparable large nonresonant contributions measured in
$B^+\to K^+K^-K^+$~\cite{Garmash:2004wa} and
$B^0\to K^+K^-K^0$~\cite{Aubert:2005kd}.
Two possible solutions for the fraction of a nonresonant background were
obtained in the first process, $(74.8 \pm 3.6)\%$ and $(65.1 \pm 5.1)\%$.
We quote the former value, which corresponds to a fit with lower $\chi^2$ (see
Table V of Ref.\ \cite{Garmash:2004wa}). These fractions and the total
branching ratios
given in Table I were used to calculate the nonresonant branching ratios. 

\begin{figure}
\begin{center}
\includegraphics
[height=3in]{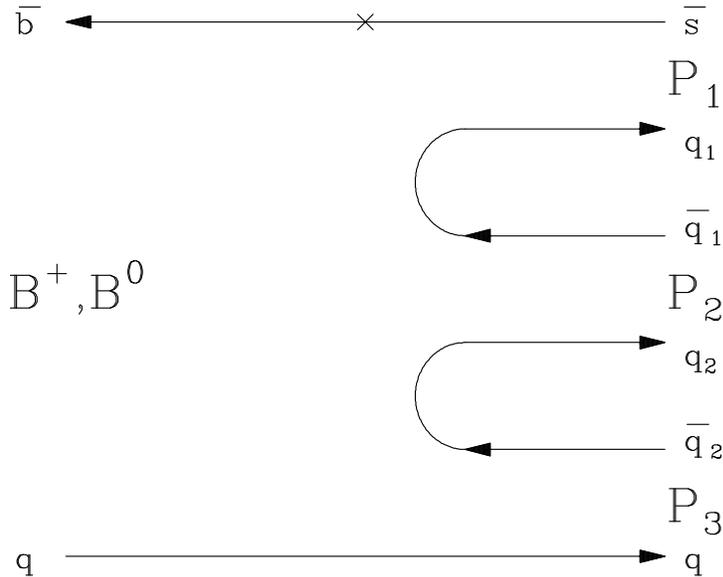}
\end{center}
\caption{Graphs describing nonresonant background in penguin-dominated 
$B \to P_1 P_2 P_3$ decays. The cross denotes a $\bar b \to
\bar s$ flavor transition.  Gluons or quarks associated with the penguin
operator are not shown explicitly.  Here $q$ denotes $u$ for a $B^+$ or
$d$ for a $B^0$.
\label{fig:peng3p}}
\end{figure}

A model describing nonresonant background amplitudes in $b\to s$ dominated 
$B\to PPP$ decays is shown in Fig.\ \ref{fig:peng3p}.  The amplitudes may be
categorized by whether the quark pairs $q_i \bar q_i~(i=1,2)$ shown in Fig.\
\ref{fig:peng3p} are $u \bar u, d \bar d$ or $s \bar s$.  Isospin symmetry is
implied by associating equal amplitudes with $u \bar u$ and $d\bar d$. This
symmetry assumption may be extended to flavor SU(3) by associating the same 
amplitude with $s \bar s$. Broken SU(3) may be represented by using a 
smaller amplitude for an $s\bar s$ pair. 

 \begin{table}
 \caption{Nonresonant background amplitudes for $B \to PPP$ decays as a 
 function of quark pairs $q_1 \bar q_1$ and $q_2 \bar q_2$.
\label{tab:amps}}
\begin{center}
\begin{tabular}{l c c c c} \hline \hline
Decaying & $q_1 \bar q_1$ & $q_2 \bar q_2$ & Final & Coefficient \\
$~~~~~B$ &                &               & state & of amplitude \\ \hline
$B^+$ & $u \bar u$ & $u \bar u$ & $K^+ \pi^0 \pi^0$ & $1/\s$ \\ 
$=\bar b u$ &      & $d \bar d$ & $K^+ \pi^- \pi^+$ & $-1$   \\
      &            & $s \bar s$ &   $K^+ K^- K^+$   & $-\s$  \\
      & $d \bar d$ & $u \bar u$ & $K^0 \pi^+ \pi^0$ & $-1/\s$ \\
      &            & $d \bar d$ & $K^0 \pi^0 \pi^+$ & $1/\s$ \\
      &            & $s \bar s$ & $K^0 \bar K^0 K^+$ &    1  \\ \hline
$B^0$ & $u \bar u$ & $u \bar u$ & $K^+ \pi^0 \pi^-$ & $1/\s$ \\
$=\bar b d$ &      & $d \bar d$ & $K^+ \pi^- \pi^0$ & $-1/\s$ \\
      &            & $s \bar s$ &   $K^+ K^- K^0$   & $-1$   \\
      & $d \bar d$ & $u \bar u$ & $K^0 \pi^+ \pi^-$ & $-1$   \\
      &            & $d \bar d$ & $K^0 \pi^0 \pi^0$ & $1/\s$ \\
      &            & $s \bar s$ & $K^0 \bar K^0 K^0$ & $\s$ \\ \hline \hline
\end{tabular}
\end{center}
\end{table}

Table \ref{tab:amps} gives the contributions to the various processes in terms 
of their coefficients.  We use conventions for states defined in Refs.\
\cite{Gronau:1994rj} and~\cite{Gronau:1995hn}.  Quark model assignments include
$B^+ = u \bar b,~B^0 = d \bar b$, with states containing a $\bar u$ quark
defined with a minus sign for convenience in isospin calculations.  Thus, a
neutral pion is $\pi^0 = (d \bar d - u \bar u)\sqrt{2}$. The entries in Table
\ref{tab:amps} contain factors of $2 \cdot 1/\s = \s$ for identical particles. 

The coefficients in Table \ref{tab:amps} imply symmetry relations between decay
rates contributed by a nonresonant background in different processes.  For
instance, the nonresonant branching ratio in $B^+\to K^+\pi^0\pi^0$ is
predicted to be half of that measured in $B^+\to K^+\pi^+\pi^-$.  Relations
applying separately to $B\to K\pi\pi$ and $B\to K\bar K K$ decays follow
from isospin symmetry and are generally expected  to hold in our model more
precisely than relations between these two types of processes which assume
flavor SU(3). Let us discuss some of these relations which can be tested using
current measurements.

An interesting prediction follows from the two equal and opposite amplitudes 
present in $B^+ \to K^0 \pi^+ \pi^0$.  When added together, the two
contributions cancel.  This is a key test of the S-wave nature (or any even 
angular momentum) of the $\pi \pi$
system for the nonresonant amplitude.  An S-wave $\pi \pi$ system with charge
$\pm 1$ must be in a state of $I=2$, which cannot be reached with the penguin
transition illustrated here.  Thus the nonresonant contributions to $B^+ \to
K^0 \pi^+ \pi^0$ and $B^0 \to K^+ \pi^- \pi^0$ are predicted to vanish if our
assumptions are valid.  The current upper bound of $4.6\times 10^{-6}$ on 
the nonresonant branching ratio of the second process is indeed much 
smaller than the other nonresonant branching ratios quoted in Table
\ref{tab:NRB}.

Table \ref{tab:amps} predicts that the nonresonant decay width for $B^+\to
K^+K^-K^+$ is two times larger than that for $B^0\to K^+K^-K^0$.  This relation
does not hold so well
(we will comment on a probable reason in the next section),
\bea\label{KKK}
\b(B^+\to K^+K^-K^+)_{\rm NRB} & = & 2(\tau_+/\tau_0)\b(B^0\to 
K^+K^-K^0)_{\rm NRB}~~; \\
22.5 \pm 1.8 & = & 37.7 \pm 4.1~~,\nonumber
\eea
where the NRB branching ratios here and subsequently are taken from
Table \ref{tab:NRB}.
This relation tests the assumption that on the right-hand-side the $K^+$ and
$K^0$ in the nonresonant background are in a symmetric $I=1$ state. In this
case the two processes involve a single isospin amplitude~\cite{Gronau:2003ep},
and their ratio of rates is given by the squared ratio of corresponding
Clebsch-Gordan coefficients, $(\s)^2 =2$. The assumption of a $K^+K^0$
symmetric state stands in contrast to the dependence on the $K^+K^-$ invariant
mass observed in the nonresonant background for $B^0\to
K^+K^-K^0$~\cite{Aubert:2005kd}.  No such dependence was observed in the
process on the left-hand-side of (\ref{KKK}).

In the SU(3) limit, one may relate the nonresonant background in the above
processes and the nonresonant amplitude in $B^+\to K^+\pi^+\pi^-$. Comparing
with $B^0\to K^+K^-K^0$, one expects
\bea\label{Kpp}
\b(B^+\to K^+\pi^+\pi^-)_{\rm NRB} & = & (\tau_+/\tau_0)\b(B^0\to 
K^+K^-K^0)_{\rm NRB}~~;
\\ 
18.4 \pm 1.9 & = & 18.8 \pm 2.0~~,\nonumber
\eea
which holds very well. Under the underlying SU(3) approximation for nonresonant
amplitudes, one would have expected this relation to be less precise than
(\ref{KKK}).  Note that in both processes appearing in (\ref{Kpp}) the measured
nonresonnt background is not exactly symmetric under interchanging the three
meson momenta, as assumed in our model.
The agreement in (\ref{Kpp}) favors the large nonresonant background in 
$B^+\to K^+\pi^+\pi^-$ given in Table \ref{tab:NRB} over the small value quoted
in Ref.~\cite{Group(HFAG):2005rb}. 

\newpage

\centerline{\bf V.  IMPLICATIONS FOR DALITZ PLOT ANALYSES}
\bigskip

The isospin relations we have quoted in the Section III are expected to be
valid separately for resonant and nonresonant contributions.  Thus, the amount
of $K^*_0(1430)^0\pi^+$ in $B^+ \to K^+ \pi^+ \pi^-$, for which a two-fold
ambiguity appears in the analysis of Ref.~\cite{Garmash:2004wa} (see Table IV
there), should be the same as the amount of $K^*_0(1430)^+\pi^-$ measured in 
$B^0 \to K^0 \pi^+ \pi^-$ [see Eq.~(\ref{1430-1})], and should be related to
the amount of $K^*_0(1430)^0\pi^0$ measured in $B^0 \to K^+ \pi^- \pi^0$ [see
Eq.~(\ref{1430-2})].  Similarly, the amount of nonresonant background in $B^+
\to K^+ \pi^+ \pi^-$, for which there seems to be some question in comparing
Refs.\ \cite{Group(HFAG):2005rb} and \cite{Abe:2005ig}, should be the same as 
in $B^0 \to K^0 \pi^+ \pi^-$ for which no value has been quoted yet.
Also, if a nonresonant background is small in $B^0 \to K^+ \pi^0 \pi^-$,
as shown in Table \ref{tab:NRB}, we would also expect it to be small in
$B^+ \to K^0 \pi^+ \pi^0$.

The isospin relation $\Gamma(B^+ \to K^+ \phi) = \Gamma(B^0 \to K^0 \phi)$
appears to be satisfied by present data.  Thus, we expect the non-$\phi$
contributions in $B \to K \bar K K$ decays, which are related to one another by
isospin reflection, also to have equal partial widths.  This is confirmed by
Eqs.~(\ref{1KKK}) and (\ref{2KKK}).

Fits to Dalitz plots often involve discrete ambiguities in assigning
amplitudes and phases to given decay channels.  This is demonstrated, for
instance, by the two largely different solutions for $\b(B^+ \to K^*_0(1430)^0
\pi^+)$ measured in Ref.~\cite{Garmash:2004wa}. Isospin symmetry, which relates
this process to $B^0\to K^*_0(1430)^+\pi^-$, resolves this
ambiguity.  Similarly, the isospin relation $\Gamma(B^+\to \chi_{c0}K^+) =
\Gamma(B^0\to \chi_{c0}K^0)$ is useful for eliminating a two-fold ambiguity in
the measurement of the left-hand-side~\cite{Garmash:2004wa}. This determines
the fraction of $\chi_{c0}K^+$ in the total of all $B^+\to K^+K^-K^+$ decays to
be about three percent rather than about eight percent, both solutions being
permitted in~\cite{Garmash:2004wa}.

One of the predictions of fully symmetric final states in nonresonant
background amplitudes is that these amplitudes should be suppressed in  
$B^+ \to K^0 \pi^+ \pi^0$ and $B^0 \to K^+ \pi^0 \pi^-$ relative to other
processes under discussion. This prediction is supported by the upper bound on
$\b(B^0\to K^+\pi^-\pi^0)_{\rm nonres}$ in Table \ref{tab:NRB}, awaiting an 
improvement in the upper bound. 

The violation of (\ref{KKK}) is probably related to the deviation from a fully
symmetric nonresonant background amplitude measured in $B^0\to
K^+K^-K^0$~\cite{Aubert:2005kd}.  The discrepancy may be the
result of the fact that nonresonant backgrounds are unstable in the fits.
This is demonstrated by the large discrepancy between two values of the
nonresonant background in $B^+\to K^+\pi^+\pi^-$ measured in Refs.\
\cite{Aubert:2004fn} and~\cite{Abe:2005ig} using different definitions for the
nonresonant background.  Another ambiguity in both $B^0\to K^+K^-K^0$ and
$B^+\to K^+K^-K^+$ is observed between a large and a very small contribution
peaking around 1500 MeV/$c^2$~\cite{Aubert:2005kd,Garmash:2004wa}.
As noted, the nonresonant background in $B^+\to K^+K^-K^+$ was found to be 
completely symmetric~\cite{Garmash:2004wa}.  Symmetry in the two identical
$K^+$ mesons follows from Bose symmetry.

\bigskip

\centerline{\bf VI.  CONCLUSIONS}
\bigskip

We have considered strangeness-changing $B \to PPP$ decays for $P = \pi, K$
under the assumption that the dominant transition is the isospin-preserving
penguin amplitude with flavor structure $\bar b \to \bar s$.  In this
approximation pairs of $B^+$ and $B^0$ decay amplitudes to $K \pi \pi$ or $K
\bar K K$ are related to one another under the isospin reflection $u \lra d$,
and final states have $I=1/2$. For decays involving more than one kaon
relations involving final states with $K_S$ and $K_L$ hold under the assumption
that $\phi K$-subtracted amplitudes are symmetric under $K \leftrightarrow \bar
K$, as measured in processes involving $K^+$ and $K^-$.

All the proposed isospin relations are obeyed experimentally where data exist,
excluding $\b(B^0\to K^{*0}\pi^0)$ which seems low relative to
$\frac{1}{2}\b(B^0\to K^{*+}\pi^-)$. The relations lead
to predictions where data are still missing.
This success led to our proposal to combine the study of Dalitz plots for 
isospin-related processes, which can resolve discrete ambiguities in fitting 
resonance parameters to given Dalitz plots.

We have presented a model for nonresonant background amplitudes in $B\to
K\pi\pi$ and $B\to K\bar KK$, which are symmetric in the three outgoing meson
momenta. Predictions characteristic to this assumption are a suppressed
nonresonant background in $B^+ \to K^0 \pi^+ \pi^0$ and $B^0 \to K^+ \pi^0
\pi^-$ and simple relations between nonresonant branching ratios in several
processes.  This approach has the potential for resolving some ambiguities in
determining nonresonant background amplitudes from fits to Dalitz plots. 

We have assumed that strangeness-changing charmless decays are dominated
by an isospin preserving $\bar b\to \bar s$ amplitude.  
These decays involve also $\Delta I=1$ electroweak penguin contributions, which
are expected to be suppressed relative to 
the dominant $I=0$ QCD-penguin amplitude~\cite{Gronau:1995hn,Fleischer:1993gr}, 
and small $\Delta I=1$ ``tree" amplitudes suppressed by $\lambda^2$ ($\lambda 
\approx 0.2$).  
The effects of these suppressed amplitudes in $B\to K\pi$ decays have been studied
recently in Refs.~\cite{Gronau:2005gz} and~\cite{Gronau:2005kz}, quoting
earlier references discussing these effects. The smallness of the effect 
is demonstrated, for instance, by the relatively small measured deviations from 
$\Delta I=0$ relations among $B\to K\pi$ decay rates. 
An example is the ratio of branching ratios~\cite{Group(HFAG):2005rb,Buras:1998rb},
\beq\label{Rn}
R^{-1}_n \equiv \frac{2\b(B^0\to K^0\pi^0)}{\b(B^0\to K^+\pi^-)} = 1.22 \pm 0.11~~,
\eeq
which differs only by $2\sigma$ from the $\Delta I=0$ value of one.

Isospin breaking effects should be considered in $B\to PPP$ when data become
sufficiently accurate.  The first case to be studied is understanding
$2\b(B^0\to K^{*0}\pi^0) < \b(B^0\to K^{*+}\pi^-)$. 
A ratio $R^{*-1}$, defined in analogy with $R^{-1}_n$,
\beq
R^{*-1}_n \equiv \frac{2\b(B^0\to K^{*0}\pi^0)}{\b(B^0\to K^{*+}\pi^-)}
 = 0.35 \pm 0.17~~,
\eeq
is $3.9\sigma$ below one, thus presenting a larger discrepancy from $\Delta I
= 0$ than measured in $R^{-1}_n$.  An interpretation for the small value of
$\b(B^0\to K^{*0}\pi^0)$ was presented in Ref.~\cite{Chiang:2003pm} in terms of
destructive interference between an electroweak penguin contribution and a QCD
penguin amplitude, the ratio of which is enhanced relative to that occurring
in $B\to K\pi$.  An interesting and pressing question is whether such
enhancement can be accounted for in the Standard Model of electroweak and
strong interactions.  While some suppression of $\b(B^0\to K^{*0}\pi^0)$ can be
accounted for in calculations based on QCD factorization~\cite{Beneke:2003zv},
central values computed for $B\to K^*\pi$ branching ratios are consistently
lower than the data by a factor 2.1 to 3.5.
Small $\Delta I =1$ contributions also may account for the $2.4 \sigma$
discrepancy in the relation (\ref{2KKK}).

We have considered CP-averaged rates, disregarding in this work possible CP 
asymmetries. The approximate relations we have derived apply separately to $B$
and $\bar B$ decays. While the $\Delta I=0$ terms in decay amplitudes are
dominated by a Cabibbo-Kobayashi-Maskawa (CKM) factor $V^*_{tb}V_{ts}$, a CKM
factor $V^*_{ub}V_{us}$ smaller by $\lambda^2$ is associated with ``tree"
contributions.  The two CKM factors involve different weak phases.  Direct CP
violation is expected if the two terms carry also different strong phases.

Potential CP asymmetries are expected from interference of decay amplitudes for
$B^+\to K^+ \chi_{c0}$, where $\chi_{c0}\to \pi^+\pi^-$ and $\chi_{c0} \to
K^+K^-$, with ``tree" amplitudes in $B^+\to K^+\pi^+\pi^-$ and $B^+\to
K^+K^+K^-$, respectively.  While a large strong phase difference is induced by
the $\chi_{c0}$ width~\cite{Eilam:1995nz},
the asymmetries depend also on the magnitudes of the smaller ``tree" amplitudes
for which calculations are model-dependent~\cite{Bajc:1998bs}. The fractions
of $K^+\chi_{c0}$ in $B^+\to K^+\pi^+\pi^-$
and $B^+\to K^+K^+K^-$ are small, at a level of three
percent~\cite{Garmash:2004wa,Aubert:2004fn} or smaller~\cite{Abe:2005ig}.

The decays $B^+\to K^+\rho^0$, where $\rho^0\to\pi^+\pi^-$, amount to a 
larger fraction of $B^+\to K^+\pi^+\pi^-$, about ten 
percent~\cite{Garmash:2004wa,Aubert:2004fn,Abe:2005ig}. 
Tentative evidence for a CP asymmetry 
in these decays has been reported recently, $A_{CP}(B^+\to K^+\rho^0) = 
0.34 \pm 0.13 \pm 0.06^{+0.15}_{-0.20}$~\cite{Aubert:2004fn},
$0.30 \pm 0.11 \pm 0.03^{+0.11}_{-0.04}$~\cite{Abe:2005ig}. 
This may be compared with a prediction, $A_{CP}(B^+\to K^+\rho^0) = 0.21 \pm
0.10$, obtained in a global SU(3) fit to all $B\to PV$
decays~\cite{Chiang:2003pm}.  A CP asymmetry at a level of 10$\%$ in the
processes discussed in this paper, resulting from penguins-tree interference as
measured in the asymmetry for $B^0\to K^+\pi^-$~\cite{Group(HFAG):2005rb}, 
would imply a small but non-negligible violation of the $\Delta I=0$ relations. 

\bigskip

\centerline{\bf ACKNOWLEDGMENTS}
\bigskip

We thank Hai-Yang Cheng, Denis Dujmic, Guy Engelhard, Guy Raz and Aaron 
Roodman for helpful discussions. This work was performed in part while M. G.
visited SLAC and while J. L. R.  was at the Aspen Center for Physics.  The
research was supported in part by the
United States Department of Energy under Grant No.\ DE FG02 90ER40560,
by the Israel Science Foundation founded by the Israel Academy of Science
and Humanities, Grant No. 1052/04, and by the German--Israeli Foundation
for Scientific Research and Development, Grant No. I-781-55.14/2003.


\begin{thebibliography}{99}

\bibitem{Gershon:2004tk}
  T.~Gershon and M.~Hazumi,
  Phys.\ Lett.\ B {\bf 596}, 163 (2004)
  [arXiv:hep-ph/0402097].

\bibitem{Aubert:2004ta}
B.~Aubert {\it et al.} [BABAR Collabortation],
  Phys.\ Rev.\ Lett.\  {\bf 93}, 181805 (2004)
  [arXiv:hep-ex/0406005].

\bibitem{Aubert:2005ja}
B.~Aubert {\it et al.} [BABAR Collaboration], 
  Phys.\ Rev.\ D {\bf 71}, 091102 (2005)
  [arXiv:hep-ex/0502019].
  
\bibitem{Aubert:2005kd}
B.~Aubert {\it et al.} [BABAR Collaboration],  
  arXiv:hep-ex/0507094.

\bibitem{Abe:2003yt}
  K.~Abe {\it et al.} [Belle Collaboration], 
  Phys.\ Rev.\ Lett.\  {\bf 91}, 261602 (2003)
  [arXiv:hep-ex/0308035].
  
\bibitem{Garmash:2003er}
A.~Garmash {\it et al.} [Belle Collaboration], 
  Phys.\ Rev.\ D {\bf 69}, 012001 (2004)
  [arXiv:hep-ex/0307082].
  
\bibitem{Chen:2005dr}
K.~F.~Chen {\it et al.}  [Belle Collaboration],  
  Phys.\ Rev.\ D {\bf 72}, 012004 (2005)
  [arXiv:hep-ex/0504023].

\bibitem{Garmash:2004wa}
 A. Garmash \ite~[Belle Collaboration], 
  Phys.\ Rev.\ D {\bf 71}, 092003 (2005)
  [arXiv:hep-ex/0412066].
  
\bibitem{Aubert:2004fn}
B.~Aubert {\it et al.} [BABAR Collaboration],    
  arXiv:hep-ex/0408032;
  arXiv:hep-ex/0507004.
  
\bibitem{Group(HFAG):2005rb}
K. Anikeev {\it et al.}, Heavy Flavor Averaging Group,
  ``Averages of b-hadron properties as of winter 2005,
  arXiv:hep-ex/0505100.
Updated results and references are tabulated periodically by this group:
{\tt http://www.slac.stanford.edu/xorg/hfag/rare.}

\bibitem{Gronau:1994rj}
  M.~Gronau, O.~F.~Hernandez, D.~London and J.~L.~Rosner,
  Phys.\ Rev.\ D {\bf 50}, 4529 (1994)
  [arXiv:hep-ph/9404283].

\bibitem{Gronau:1995hn}
  M.~Gronau, O.~F.~Hernandez, D.~London and J.~L.~Rosner,
  Phys.\ Rev.\ D {\bf 52}, 6374 (1995)
  [arXiv:hep-ph/9504327].
 
\bibitem{Chiang:2004nm}
  C.~W.~Chiang, M.~Gronau, J.~L.~Rosner and D.~A.~Suprun,
  Phys.\ Rev.\ D {\bf 70}, 034020 (2004)
  [arXiv:hep-ph/0404073].

\bibitem{Chiang:2003pm}
  C.~W.~Chiang, M.~Gronau, Z.~Luo, J.~L.~Rosner and D.~A.~Suprun,
  Phys.\ Rev.\ D {\bf 69}, 034001 (2004)
  [arXiv:hep-ph/0307395].

\bibitem{DGR}
A.~S.~Dighe, M.~Gronau and J.~L.~Rosner,
Phys.\ Lett.\ B {\bf 367}, 357 (1996) [arXiv:hep-ph/9509428]; {\it Erratum-ibid.} B
{\bf 377}, 325 (1996);
Phys.\ Rev.\ Lett.\ {\bf 79}, 4333 (1997) [arXiv:hep-ph/9707521];
  C.~W.~Chiang, M.~Gronau and J.~L.~Rosner,
  Phys.\ Rev.\ D {\bf 68}, 074012 (2003)
  [arXiv:hep-ph/0306021].

\bibitem{Eilam:1995nz}
  G.~Eilam, M.~Gronau and R.~R.~Mendel,
  Phys.\ Rev.\ Lett.\  {\bf 74}, 4984 (1995)
  [arXiv:hep-ph/9502293].
 
\bibitem{Bajc:1998bs}
  B.~Bajc, S.~Fajfer, R.~J.~Oakes, T.~N.~Pham and S.~Prelovsek,
  Phys.\ Lett.\ B {\bf 447}, 313 (1999)
  [arXiv:hep-ph/9809262];
  S.~Fajfer, R.~J.~Oakes and T.~N.~Pham,
  Phys.\ Rev.\ D {\bf 60}, 054029 (1999)
  [arXiv:hep-ph/9812313];
  S.~Fajfer, R.~J.~Oakes and T.~N.~Pham,
  Phys.\ Lett.\ B {\bf 539}, 67 (2002)
  [arXiv:hep-ph/0203072];
  S.~Fajfer, T.~N.~Pham and A.~Prapotnik,
  Phys.\ Rev.\ D {\bf 70}, 034033 (2004)
  [arXiv:hep-ph/0405065].

\bibitem{Cheng:2002qu}
  H.~Y.~Cheng and K.~C.~Yang,
  Phys.\ Rev.\ D {\bf 66}, 054015 (2002)
  [arXiv:hep-ph/0205133];
  H.~Y.~Cheng, C.~K.~Chua and A.~Soni,
  arXiv:hep-ph/0506268;
  H.~Y.~Cheng, C.~K.~Chua and K.~C.~Yang,
  arXiv:hep-ph/0508104.
  
\bibitem{Minkowski:2004xf}
  P.~Minkowski and W.~Ochs,
  Eur.\ Phys.\ J.\ C {\bf 39}, 71 (2005)
  [arXiv:hep-ph/0404194].

\bibitem{Gronau:2003ep}
  M.~Gronau and J.~L.~Rosner,
  Phys.\ Lett.\ B {\bf 564}, 90 (2003)
  [arXiv:hep-ph/0304178].

\bibitem{Grossman:2003qp}
  Y.~Grossman, Z.~Ligeti, Y.~Nir and H.~Quinn,
  Phys.\ Rev.\ D {\bf 68}, 015004 (2003)
  [arXiv:hep-ph/0303171];
  G.~Engelhard, Y.~Nir and G.~Raz,
  arXiv:hep-ph/0505194;
  G.~Engelhard and G.~Raz,
  arXiv:hep-ph/0508046.

\bibitem{Abe:2005kr}
  K.~Abe {\it et al.} [Belle Collaboration],
  arXiv:hep-ex/0509047.

\bibitem{Aubert:2005id}
B.~Aubert {\it et al.} [BABAR Collaboration], 
  arXiv:hep-ex/0508017.

\bibitem{Eidelman:2004wy}
  S.~Eidelman {\it et al.}  [Particle Data Group],
  Phys.\ Lett.\ B {\bf 592}, 1 (2004).
  
\bibitem{Abe:2005ig}
 K.~Abe \ite~[Belle Collaboration],
  arXiv:hep-ex/0509001.

\bibitem{Fleischer:1993gr}
  R.~Fleischer,
  Z.\ Phys.\ C {\bf 62}, 81 (1994).
  
\bibitem{Gronau:2005gz}
  M.~Gronau and J.~L.~Rosner,
  Phys.\ Rev.\ D {\bf 71}, 074019 (2005)
  [arXiv:hep-ph/0503131].

\bibitem{Gronau:2005kz}
  M.~Gronau,
  Phys.\ Lett.\ B {\bf 627}, 82 (2005).
  
\bibitem{Buras:1998rb}
  A.~J.~Buras and R.~Fleischer,
  Eur.\ Phys.\ J.\ C {\bf 11}, 93 (1999)
  [arXiv:hep-ph/9810260].
  
\bibitem{Beneke:2003zv}
  M.~Beneke and M.~Neubert,
  Nucl.\ Phys.\ B {\bf 675}, 333 (2003)
  [arXiv:hep-ph/0308039].

\end{thebibliography}
\end{document}